\documentclass[preprint,aps ,nofootinbib]{revtex4}
\usepackage{graphicx}
\usepackage{amsmath}
\usepackage{amsfonts}
\usepackage{amssymb}
\usepackage{color}%
\usepackage{dcolumn}
\newcommand{\f}{\begin{equation}}
\newcommand{\ff}{\end{equation}}
\newcommand{\fa}{\begin{eqnarray}}
\newcommand{\ffa}{\end{eqnarray}}

\begin{document}
\title{Chaos from the ring string in Gauss-Bonnet black hole in $AdS_5$ space}
\author{Da-Zhu Ma $^{1}$} \email{mdzhbmy@126.com}
\author{Jian-Pin Wu $^{2}$} \email{jianpinwu@gmail.com}
\author{Jifang Zhang $^{3}$} \email{jifangzhangbh@gmail.com}
\affiliation{ $^1$Department of Physics, School of Science, Hubei Minzu University, Enshi, 445000, China\\
$^2$Department of Physics, School of Mathematics and Physics, Bohai University, Jinzhou, 121013, China\\
$^3$Teaching and research institute of college computer, Bohai University, Jinzhou, 121013, China}

\begin{abstract}

By the Poincare sections, the dynamical behaviors of the ring string in the AdS-Gauss-Bonnet (AdS-GB) black hole are studied.
A threshold value of the Gauss-Bonnet (GB) parameter $\lambda$ are found by the numerical method, below which the behavior of this dynamical system is
no-chaotic and above which the behavior becomes gradually chaotic.
The chaotic behavior becomes stronger with the increases of GB parameter.
It is different from the case in the AdS-Schwarzschild black hole, which is weakly chaotic.
Furthermore, we confirm our findings by the Fast Lyapunov indicators.

\end{abstract} \maketitle

\section {Introduction}

Chaos theory studies the behavior of dynamical systems that are highly sensitive to initial conditions and has been applied in general relativity and cosmology.
Due to the integrability of the point particle motion in the generic Kerr-Newman background\cite{Carter:1968},
to have chaotic point particle dynamics, we have to study some complicated multi black hole geometries such as the Majumdar-Papapetrou geometry\cite{Majumdar:1947,Papapetrou:1947},
in which the chaotic behavior has been explored in\cite{gr-qc/9402027,gr-qc/0610119}, or charged particles in a magnetic field
interacting with gravitational waves\cite{Varvoglis:1992}, or particles near a black hole in a Melvin magnetic
universe\cite{Karas:1992}, or in a perturbed Schwarzschild spacetime\cite{Bombelli:1992,gr-qc/9604032,gr-qc/9505036}.

Therefore, to have a chaotic dynamics in some simple and symmetric systems,
the ring string can be introduced instead of point particle.
Some pioneer works have been done to study the chaotic dynamics along this direction.
As for the Schwarzschild black hole in asymptotically flat space, it was shown that the ring string dynamics is chaotic\cite{gr-qc/9908039}.
Furthermore, the chaotic behavior of the ring string is also found in AdS-Schwarzschild black hole\cite{1007.0277},
in which they give some concrete evidence supporting their finding in terms of the power spectrum, the largest Lyapunov exponent, Poincare sections and basins of attractions.
In particular, they propose that the operators being described by the ring string are generalizations of the Gubser-Klebanov-Polyakov\cite{hep-th/0204051} operators
and try to extend the AdS/CFT dictionary to include the chaotic behavior, in which they propose that a positive largest Lyapunov exponent on the gravity side
corresponds to the appropriate bound for the time scale of Poincare recurrences on the gauge theory side.
In addition, there are a number of recent works developing the topic of chaos and nonintegrability\cite{1201.5634,1311.3241,1311.1521,1211.3727,1105.2540,1103.4101,1103.4107} and exploring the role of quantum chaos\cite{1304.6348,1209.5902} in the AdS/CFT correspondence.

On the other hand, in terms of the point of view of the AdS/CFT, the higher curvature interactions on the bulk are identified with the finite
coupling corrections on the dual field theory. Therefore, it is interesting and necessary to study the higher
curvature interactions in an effective gravitational theory before string theory is fully understood
such that we can broaden the class of the dual field theories one can holographically study.
A simple and useful effective gravitational theory including higher curvature interactions is Gauss-Bonnet (GB) gravity,
which includes only the curvature-squared interaction. Several holographic models, including the holographic superconductors and holographic fermions,
have been explored in this set up\cite{0907.3203,0912.2475,1005.4743,1007.3321,1009.1991,1002.4901,1003.5130,1103.3982,1205.6674}.
In addition, some holographic studies in the higher curvature interactions theory of gravity, including quasitopological gravity, the gravity theory with Weyl corrections, were also exploited in Refs. \cite{1003.4773,1003.5357,1008.4066,1010.0700,0811.4195,1010.0443,1010.1929,PLBWeylMa}.
All the above studies can give some interesting and significant results from higher curvature interactions.
So, it is interesting to extend the explorations of the study on the dynamical behavior of the ring string in AdS-Schwarzschild black hole in\cite{1007.0277}
to that in AdS-GB black hole. It is just our subject in this paper.

Our paper is organized as follows. In section \ref{SectionGB}, we present a brief review on GB Black Holes in AdS space.
The equations of motion in phase space are derived in section \ref{SectionEOMGB}.
We introduce the numerical methods used in this paper in section \ref{SectionN} and the corresponding
chaos indicators, including Poincare sections and Lyapunov indicator, are calculated numerically in section \ref{SectionCI}.
Finally, the conclusions and discussions follow in section \ref{SectionC}.

\section {Gauss-Bonnet Black Holes in AdS space}\label{SectionGB}

In this section, we will give a brief review on GB Black Holes in AdS space.
Conventionally, we consider the following Einstein-Gauss-Bonnet action\cite{GBBoulware:1985,hep-th/0109133}:
\begin{eqnarray}
\label{GBaction}
S=\frac{1}{16\pi G_5}\int d^{5}x \sqrt{-G}\left[R+\frac{12}{L^{2}}+\lambda\left(R^{abcd}R_{abcd}-4R^{ab}R_{ab}+R^{2}\right)\right],
\end{eqnarray}
where $\lambda$ is GB coupling constant. $L$ is the AdS radius and for convenience, we will set $L=1$ in what follows.
Applying the principle of variation to the above action (\ref{GBaction}), one can easily obtain the equations of motion:
\begin{eqnarray}
\label{EinsteinE}
R_{ab}-\frac{1}{2}Rg_{ab}+6g_{ab}-2\lambda\left[H_{ab}-\frac{1}{4}Hg_{ab}\right]
=0,
\end{eqnarray}
where
\begin{eqnarray}
\label{TensorHab}
H_{ab}=R_{a}^{cde}R_{bcde}-2R_{ac}R_{b}^{c}-2R_{acbd}R^{cd}+RR_{ab},
\end{eqnarray}
and
\begin{eqnarray}
\label{TensorHtrace}
H=H_{a}^{a}.
\end{eqnarray}
The above equation has the static spherically symmetric solution as follows
\begin{eqnarray}
\label{MetricA}
ds^{2}=-f(r)dt^{2}+\frac{dr^{2}}{f(r)}+r^{2}d\Omega_3^2,
\end{eqnarray}
in which
\begin{eqnarray}
\label{metricf}
f(r)=k+\frac{r^2}{4\lambda}-\frac{r^2}{4\lambda}\sqrt{1+\frac{8\lambda M}{r^4}-8\lambda},
\end{eqnarray}
where $M$ denotes the mass of the black hole and $k=1, 0, -1$ corresponds to the sphere, plane and hyperbola symmetric cases, respectively.
In this paper, we only focus on the sphere symmetric case, i.e. $k=1$,
in which
\begin{eqnarray}
\label{metricOmega}
d\Omega_3^2=d\theta^2+\sin^2\theta d\psi^2+\cos^2\theta d\phi^2.
\end{eqnarray}
Near the boundary ($r\rightarrow \infty$), the redshift factor becomes
\begin{eqnarray}
f(r)|_{r\rightarrow\infty}=1+\frac{r^2}{4\lambda}(1-\sqrt{1-8\lambda}).
\end{eqnarray}
So, to have a well-defined Anti-de Sitter vacuum for the gravity theory, the condition $\lambda\leq 1/8$ should be imposed.

\section{Ring string in Gauss-Bonnet black holes}\label{SectionEOMGB}

To study the chaotic behaviors of the ring string in the GB black hole,
we begin with the following Polyakov action,
\begin{eqnarray}\label{Polyakovaction}
\mathcal{L}=-\frac{1}{2\pi \alpha'}\sqrt{-g}g^{\mu\nu}G_{ab}\partial_\mu X^a\partial_\nu X^b,
\end{eqnarray}
which is on-shell equivalent the Nambu-Goto action.
Here $G_{ab}$ is the spacetime metric and $X^a$ denote the coordinates of the string.
$g_{\mu\nu}$ is the induce metric on the worldsheet of the string,
which is parameterized by the coordinates $\sigma^\mu=(\tau, \sigma)$.
$\alpha'$ relates the string length $l_s$ by $l_s^2=\alpha'$.

Following the Ref.\cite{1007.0277}, we consider the following embedding
\begin{eqnarray}\label{Embedding}
t=t(\tau),~~r=r(\tau),~~\theta=\theta(\tau),~~\phi=\phi(\tau),~~\psi=\alpha\sigma,
\end{eqnarray}
where $\alpha$ plays the role of winding, which denotes the differences between strings and particles.
Here we will work in the conformal gauge, i.e., $g_{\mu\nu}=\eta_{\mu\nu}$.
And then, in terms of the metric Eqs.(\ref{MetricA}), (\ref{metricf}) and (\ref{metricOmega}),
the Polyakov Lagrangian (\ref{Polyakovaction}) takes the following concrete form
\begin{eqnarray}\label{PolyakovactionConcrete}
\mathcal{L}=-\frac{1}{2\pi \alpha'}\left[f\dot{t}^2-\frac{\dot{r}^2}{f}-r^2(\dot{\theta}^2+\cos^2\theta\dot{\phi}^2)+r^2\alpha^2\sin^2\theta\right],
\end{eqnarray}
where the dot represents the derivative with respect $\tau$ (the same hereinafter) and
we will denote the derivative with respect to $r$ by using prime in what follows.
Using the canonical transform, the corresponding Hamiltonian can be rewritten as
\begin{eqnarray}\label{Hamiltonian}
H = \frac{\pi\alpha'}{2}\bigg[fp_r^2+\frac{p_{\theta}^2}{r^2}+\frac{p_{\phi}^2}{r^2\cos^2\theta}-\frac{p_t^2}{f}\bigg]
+\frac{1}{2\pi \alpha'}\,r^2\alpha^2 \sin^2\theta,
\end{eqnarray}
where $\{t,~p_t\}$, $\{r,~p_r\}$, $\{\theta,~p_\theta\}$ as well as $\{\phi,~p_\phi\}$ are the canonical phase space variables and
\begin{eqnarray}\label{CMomentum}
p_t= -\frac{f\dot{t}}{\pi \alpha'}, \quad p_r= \frac{\dot{r}}{\pi \alpha' f}, \quad
p_{\theta}=\frac{r^2\dot{\theta}}{\pi \alpha'}, \quad p_{\phi}= \frac{r^2\cos^2\theta\, \dot{\phi}}{\pi \alpha'}.
\end{eqnarray}
Then, the canonical equations of motion can be calculated as following
\fa
&&
\label{tEOM}
\dot{t}:=\{t,H\}=-\frac{\pi \alpha'}{f}p_t,
\\
&&
\label{ptEOM}
\dot{p}_t:=\{p_t,H\}=0,
\\
&&
\label{rEOM}
\dot{r}:=\{r,H\}=\pi \alpha' f p_r,
\\
&&
\label{prEOM}
\dot{p_r}:=\{p_r,H\}=-\frac{\pi\alpha'f'}{2f^2}p_t^2 -\frac{\pi\alpha' f'}{2} p_r^2 +\frac{\pi\alpha'}{r^3}p_{\theta}^2 +\frac{\pi\alpha'}{r^3\, \cos^2\theta}p_{\phi}^2
- \frac{\alpha^2 r \sin^2\theta}{\pi\alpha'},
\\
&&
\label{thetaEOM}
\dot{\theta}:=\{\theta,H\}= \frac{\pi \alpha'}{r^2}p_\theta,
\\
&&
\label{pthetaEOM}
\dot{p_\theta}:=\{p_\theta,H\}=-\frac{\pi\alpha'\sin\theta}{r^2\cos^3\theta}p_\phi^2-\phi\alpha'\alpha^2 r^2 \sin\theta \cos\theta,
\\
&&
\label{phiEOM}
\dot{\phi}:=\{\phi,H\}=\frac{\pi \alpha'}{r^2\cos^2\theta}p_\phi,
\\
&&
\label{pphiEOM}
\dot{p_\phi}:=\{p_\phi,H\}=0.
\ffa
From Eqs.(\ref{ptEOM}) and (\ref{pphiEOM}), we have two constants of motion $p_t=E$ and $p_\phi=l$, which relate to the energy and angular momentum, respectively.
In terms of these two constants of motion, we have the following reduced Hamiltonian system
\fa
&&
\label{H}
H =\frac{\pi \alpha'}{2}\bigg[f\,p_r^2+\frac{1}{r^2}p_\theta^2 +\frac{l^2}{r^2\cos^2\theta}-\frac{E^2}{f}\bigg]+\frac{1}{2\pi \alpha'} \alpha^2\,r^2 \sin^2\theta,
\\
&&
\label{Hr}
\dot{r}=\pi \alpha' \,\, f\,p_r,
\\
&&
\label{Hpr}
\dot{p_r}=-\frac{\pi\alpha'f'}{2f^2}E^2-\frac{\pi\alpha'f'}{2}\,p_r^2 +\frac{\pi\alpha'}{r^3}p_\theta^2 +\frac{\pi\alpha'l^2}{r^3\cos^2\theta}-\frac{\alpha^2 r \sin^2\theta}{\pi\alpha'},
\\
&&
\label{Htheta}
\dot{\theta}= \frac{\pi\alpha'}{r^2}p_\theta,
\\
&&
\label{Hp}
\dot{p}_\theta= -\frac{\pi\alpha'l^2\sin\theta}{r^2\cos^3\theta}-\frac{\alpha^2 r^2 \sin\theta\cos\theta}{\pi\alpha'}.
\ffa
In addition, we note that the Hamilton $H$ satisfy the constraint $H=0$, which follows from fixing the conformal gauge
\fa
G_{ab}\left(\partial_\tau X^a\partial_\tau X^b+\partial_\sigma X^a\partial_\sigma X^b\right)=0.
\ffa

Before proceeding, we will discuss the conserved quantities of motion in the bulk,
which correspond to the quantum numbers of the operator in the dual field theory.
Usually, the conjugate momentum can be calculate by
\begin{eqnarray}\label{Conjugate}
P^i_a=-\frac{1}{2\pi \alpha'}G_{ab}\partial^i X^b,~~~~i=\tau,~\sigma.
\end{eqnarray}
So, we have
\fa
&&
\label{P1}
P_t^\tau=\frac{E}{2\pi\alpha'},
\\
&&
\label{P2}
P_\phi^\tau=\frac{l}{2\pi\alpha'},
\\
&&
\label{P3}
P_\psi^\sigma=-\frac{\alpha r^2\sin^2\theta}{2\pi\alpha'}=-\frac{\alpha R^2}{2\pi\alpha'},
\ffa
In the third equation above, we have defined $R:=r^2\sin^2\theta$, which is the radius of the ring string.

\section{Numerical method}\label{SectionN}

It is well known that a reliable numerical method is the basic tool for investigating nonlinear dynamics,
because numerical errors may produce pseudo chaotic behavior.
Within the last decade or so, two independent groups have been engaged in the study of controlling numerical errors.
One uses the traditional numerical methods such as Runge-Kutta algorithms.
However, the low-order methods are full of artificial dissipation and
the high-order methods are time-consuming.
Especially, they both cannot conserve constancy of any first integrals for long time integration.
The other method pays attention to symplectic integrator which is widely used in conservative Hamiltonian system.
But it is limited to its application because a given difference scheme should be kept canonical while constructing higher-order symplectic algorithms.

\begin{figure}
\center{
\scalebox{0.8}[0.8]{\includegraphics {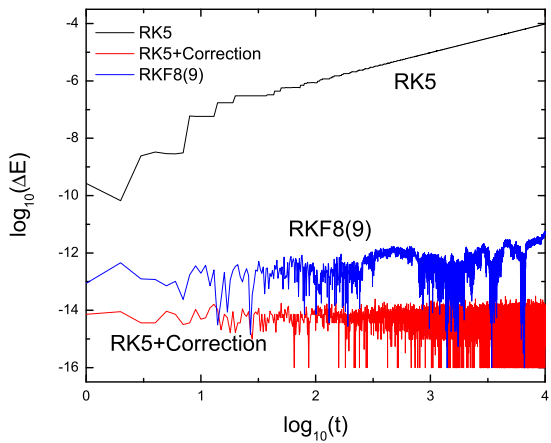}}
\caption{\label{energy}Errors of energy with time computed by RK5, RK8(9) and the velocity correction method(RK5+Correction).}}
\end{figure}

Our aim is to detect chaos in such system in this paper.
So we are particularly interested in the numerical effectiveness firstly.
To solve the Hamiltonian system of Eqs.(\ref{H})-(\ref{Hp}), a Fifth-order Runge-Kutta algorithm(RK5) is adopted as the basic integrator.
As is shown in FIG.\ref{energy}, errors of energy for RK5 have nearly the linear growth with time,
it means that the integrated coordinates $(r, \theta)$ and momentums $(p_{r}, p_{\theta})$ have deviated the original hypersurface.
This will be an adverse effect on the study of chaos later.
However, as a pioneering work, the rigorous velocity correction method of energy \cite{AJWu,NewAMa,IJMPCMa,APJMa} is a
good idea to pull the deviated hypersurface back in a least-squares shortest path.
Some numerical examples in the references have shown that the accuracy of numerical solution can be improved greatly by frequently adjustment.
Due to the energy of the system is subjected to the constraint $H=0$, it could be treated as a conserved quantity.
In this sense, the applicability of velocity correction methods of energy can be implemented here.
A dimensionless parameter $\zeta$ is used to adjust between the numerical solution $(p_{r}, p_{\theta})$ and the true value $(p^{*}_{r}, p^{*}_{\theta})$ in the form of
\begin{eqnarray}
\label{metricfv1}
P^{*}_{r}=\zeta \cdot P_{r}, ~~~~  P^{*}_{\theta}=\zeta \cdot P_{\theta}.
\end{eqnarray}
Substituting Eq.(\ref{metricfv1}) into Eqs.(\ref{H})-(\ref{Hp}), we can easily determine $\zeta$
\begin{eqnarray}
\label{metricfv2}
\zeta=\sqrt{\frac{\frac{2(H^{*}-\frac{\alpha^{2}r^{2}sin^{2}\theta}{2\pi\alpha'})}{\pi\alpha'}-\frac{l^{2}}{r^{2}cos^{2}\theta}+\frac{E^{2}}{f}}{fp^{2}_{r}+\frac{p^{2}_{\theta}}{r^{2}}}},
\end{eqnarray}
where $H^{*}=0$. Because it provides great control on energy error when it is employed,
the precision of the energy in the system of Eqs.(\ref{H})-(\ref{Hp}) at every integration step can hold perfectly which is shown in Fig.1.
It is easy to see that the method makes the energy error almost reach to the double precision of machine at every integration step,
and it is better than RKF8(9) which is used to compare.
It displays sufficiently that this method is very powerful so that it can avoid the pseudo chaos caused by numerical errors.

\section{Chaos indicators}\label{SectionCI}

Up to now, it still has no accurate definition of chaos.
We consider chaos is a kind of seeming random,
chance or irregular movement, which appears in a definiteness system.
The most important characteristics of chaos is that it is sensitive to initial value.
It should be pointed out that there are two factors usually affect detection of chaos.
One is the numerical algorithms which we have discussed before. The other factor is chaos indicators.
There are many kinds of chaos indicators, such as the Poincare surfaces of section,
the Lyapunov characteristic exponents (LCEs),
the maximal LCE, the Fast Lyapunov indicators (FLI), the method of fractal basin boundaries and so on.
Each of the known chaos indicators has its advantages and disadvantages.
Since the system of Eqs.(\ref{H})-(\ref{Hp}) is a three-dimensional manifold,
the Poincare surfaces of section, which is regarded as a topologically invariant procedure, is enough to reveal the phase space structure.
Lyapunov Indictors are often used in higher dimensional space,
they also can be used to check the accuracy of the Poincare section method.
Especially, the Fast Lyapunov indicators (FLI) is more effective and faster to reveal chaos orbital than another kinds of lyapunov indicators.
So we only focus on this two Chaos indicators in the paper.
In the next subsection we will study the effects of the GB parameter on chaos by calculating numerically the Poincare sections and the Fast Lyapunov indicators.

\subsection{Poincare sections}

It should be stressed that the choice of orbit is arbitrary in a deterministic system.
For the system of Eqs.(\ref{H})-(\ref{Hp}), we only require the initial value lies in interior of the bounded system.
Here, we will set $\alpha'=1/\pi$, $\beta=1$ and $r_H=1$.
$r_H$ is the horizon radius and satisfies $f(r_H)=0$, by which, we can determine the parameter $M$.
In our simulations, the initial conditions of $r$, $\theta$ and $p_{r}$ are given in
admissible regions of the system of Eqs.(\ref{H})-(\ref{Hp}), while the starting value of $p_{\theta}$($p_{\theta} >0$) can be solved from Eq.(\ref{H})
with the energy constraint $H=0$ and a given value of $\lambda$.
The stable region can also be found numerically if the GB parameter $\lambda$ is fixed.
We take $r_{0}=5$ and $\theta_{0}=0$ for example, some stable regions are given in TABLE\ref{Tabeldeltavsp}.
\begin{widetext}
\begin{table}[ht]
\begin{center}
\begin{tabular}{|c|c|c|c|c|c|c|c|c|c|}
         \hline
~$\lambda$~ &~$-0.1$~&~$-0.05$~&~$0.01$~&~$0.05$~
          \\
        \hline
~$p_{r0}$~ & ~$[2.42,5.17]$~ & ~$[2.67,5.05]$~ & ~$[2.90,4.91]$~&~$[3.07,4.82]$~
          \\
        \hline
\end{tabular}
\caption{\label{Tabeldeltavsp}The stable zones for $p_{r0}$ with different $\lambda$.
}
\end{center}
\end{table}
\end{widetext}
Chaotic behavior occurs if $p_{r0}$ is outside the stable region, any track in the unstable region may be chaotic.
Periodic orbit region narrows down as the GB parameter increased.
In other words, chaotic behavior is more likely to happen in larger GB parameter value.

\begin{figure}
\center{
\scalebox{0.8}[0.8]{\includegraphics{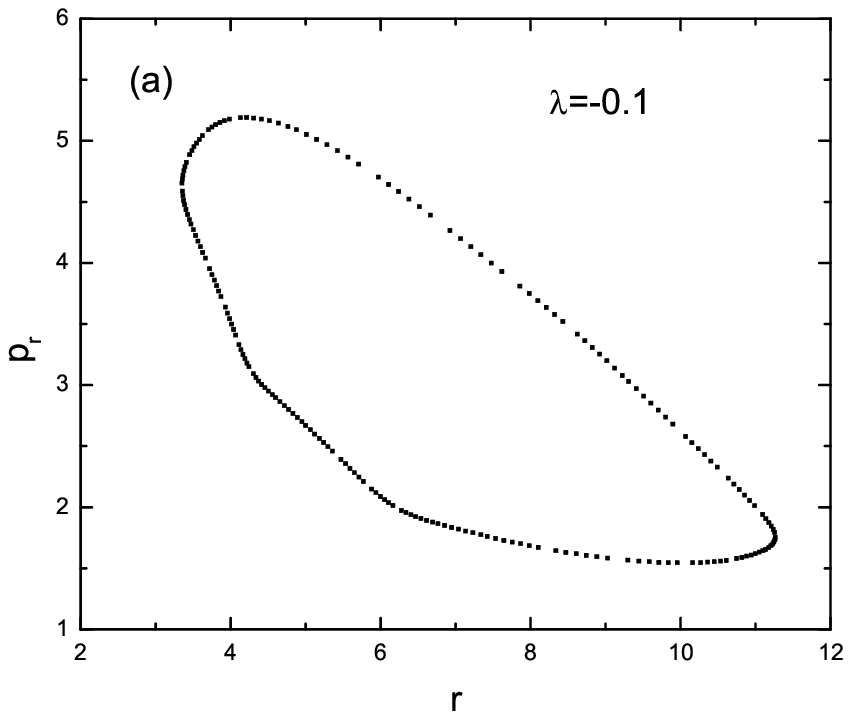}}
\scalebox{0.8}[0.8]{\includegraphics{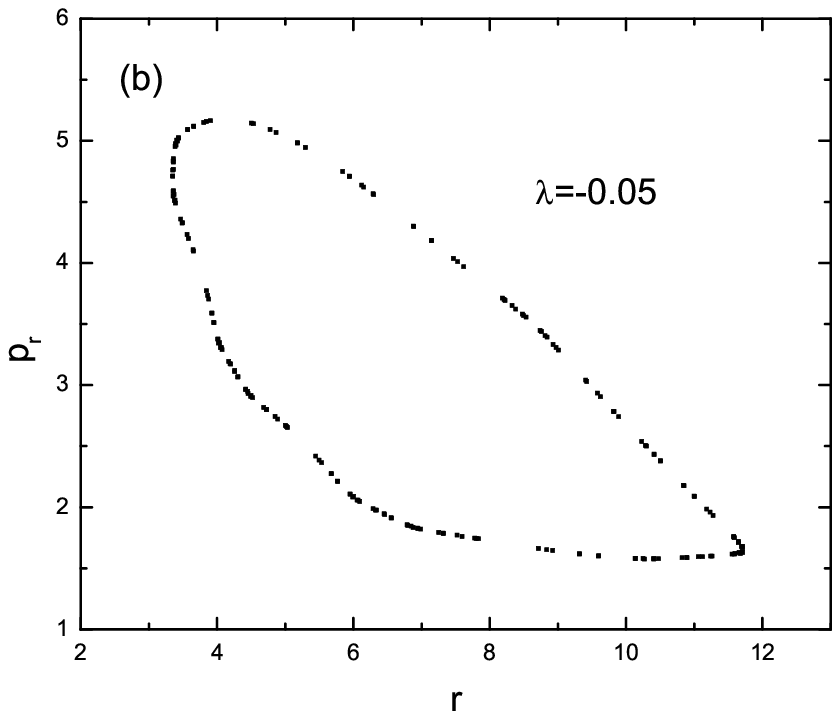}}
\scalebox{0.8}[0.8]{\includegraphics{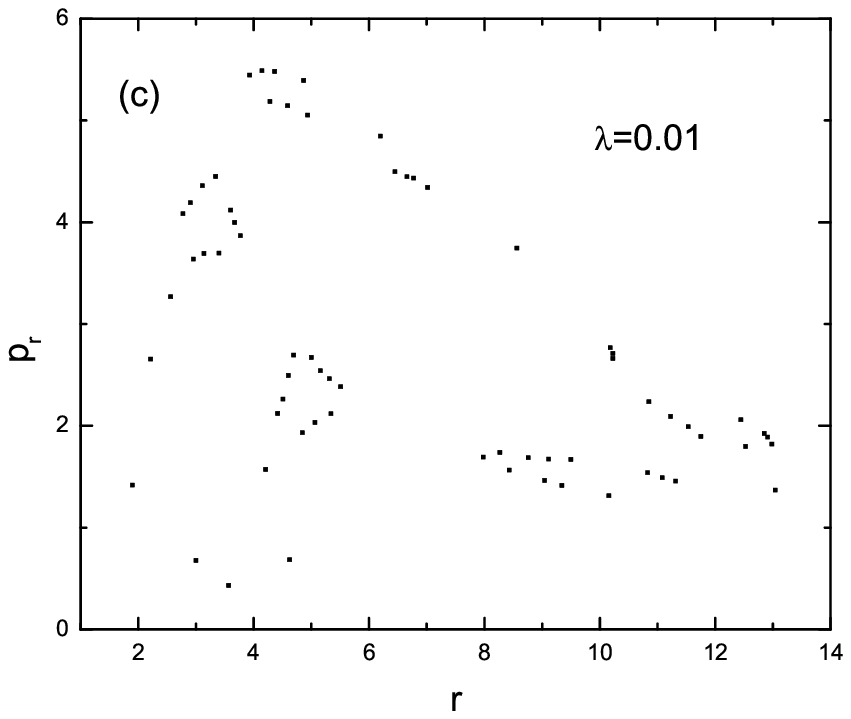}}
\scalebox{0.8}[0.8]{\includegraphics{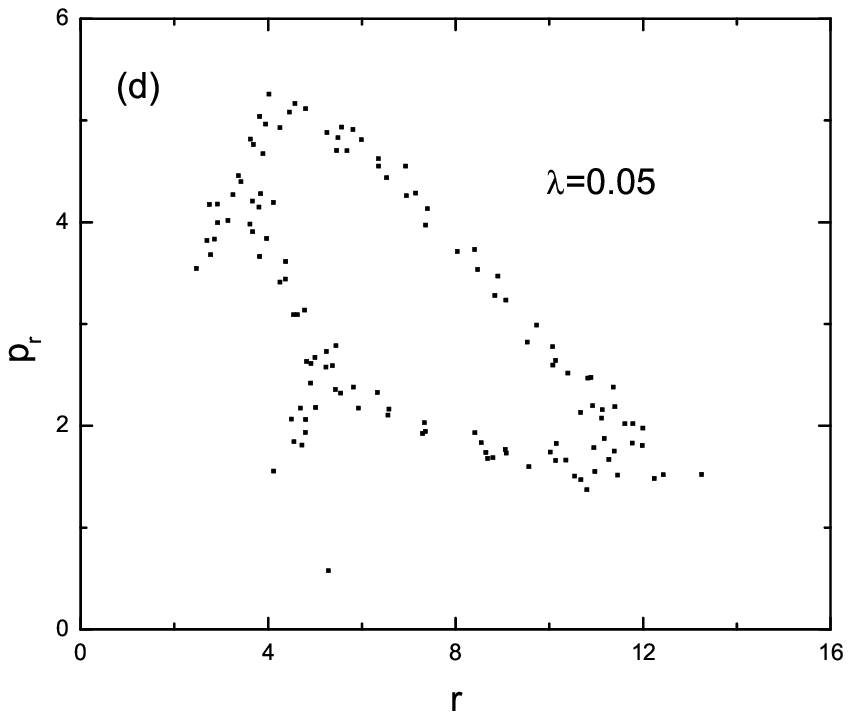}}
\caption{\label{PS}Poincare sections on the plane ($\theta=0, p_{\theta}>0$) with $H=0$, $r_{0}=5$, $p_{r0}=2.670855$ and $\theta_{0}=0$.}}
\end{figure}

Without loss of generality, we shall choose the same initial conditions as that in Ref.\cite{1007.0277}, i.e., $r_{0}=5$, $p_{r0}=2.670855$ and $\theta_{0}=0$, in what follows.
For this fixed initial values, we found numerically a threshold value for $\lambda$($\lambda\simeq-0.05$),
below which the behavior of this dynamical system is no-chaotic and above which the behavior becomes gradually chaotic.
The results are showed in FIG.\ref{PS}. From FIG.\ref{PS}(a) and (b), we can see that the trajectory is a quasi-periodic Kolmogorov-Arnold-Moser (KAM) tori,
which indicates the behavior of this system for $\lambda=-0.1$ and $\lambda=-0.05$ is non-chaotic.
However, with the increases of the GB parameter $\lambda$ (beyond $-0.05$),
the behavior of this system changes from non-chaotic to chaotic (FIG.\ref{PS}(c) and (d))
in which the KAM tori is destroyed. Especially for $\lambda=0.05$ in FIG.\ref{PS}(d),
we discovered that the tori is completely composed of discrete points.
It indicates that the chaotic behavior is stronger with the increases of the GB parameter $\lambda$.
It should be noted that the dynamical behavior of the ring string on the AdS-Schwarzschild geometry (corresponding to $\lambda=0$ here) is weakly chaotic\cite{1007.0277},
which is consistent with our finding here.

To validate the accuracy of our result, we will calculate numerically the Fast Lyapunov indicators (FLI)
in the next subsection.
\subsection{Fast Lyapunov indicator}

The Lyapunov indicator is often adopted to measure two adjacent orbitals over time with the average separation index,
which can be used to determine characteristics (periodic or chaotic) of the orbit.
There are two ways to calculate the Lyapunov exponent in the references.
One is the variational method, the other is the two-particle method \cite{AJTancredi}.
In general, the latter is more convenient in application.
However, there are three negative factors should be pointed out:
$(1)$ It is hard to choose the initial distances of the two nearby orbitals in the configuration space.
$(2)$ It is difficult to get an appropriate renormalization time.
$(3)$ Both will cost much long time to achieve a stabilizing limit value of the Lyapunov exponent in chaotic track.
In contrast, a quicker and more sensitive indicator to find chaos is the Fast Lyapunov indicator (FLI).

\begin{figure}
\center{
\scalebox{0.8}[0.8]{\includegraphics{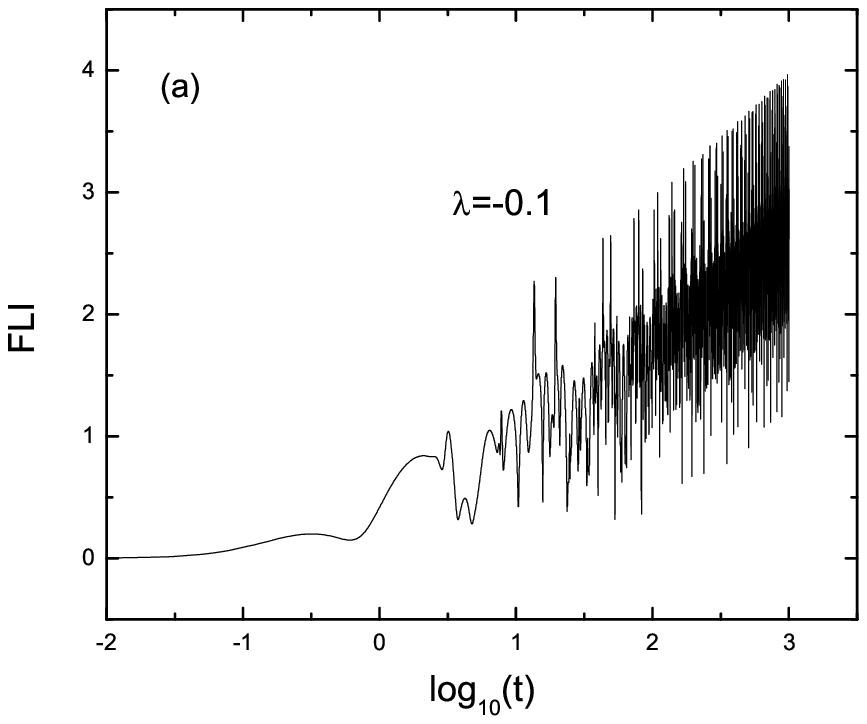}}
\scalebox{0.8}[0.8]{\includegraphics{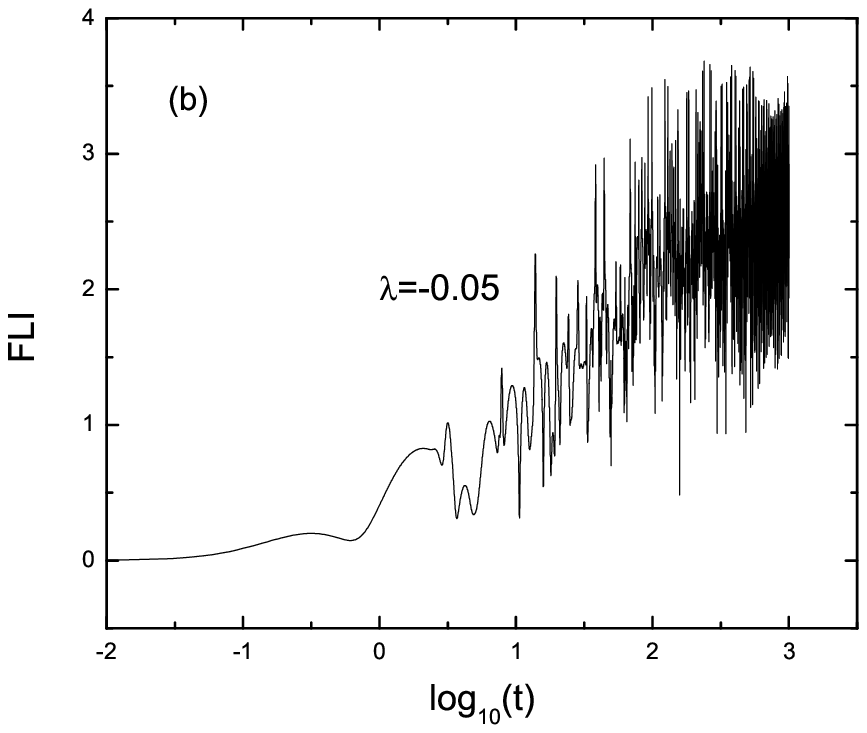}}
\scalebox{0.8}[0.8]{\includegraphics{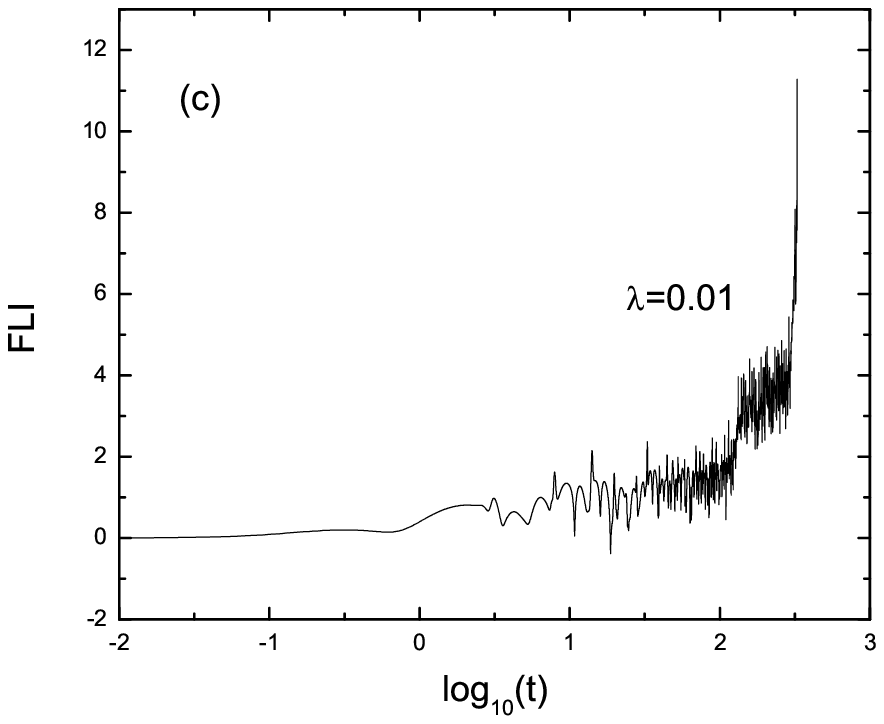}}
\scalebox{0.8}[0.8]{\includegraphics{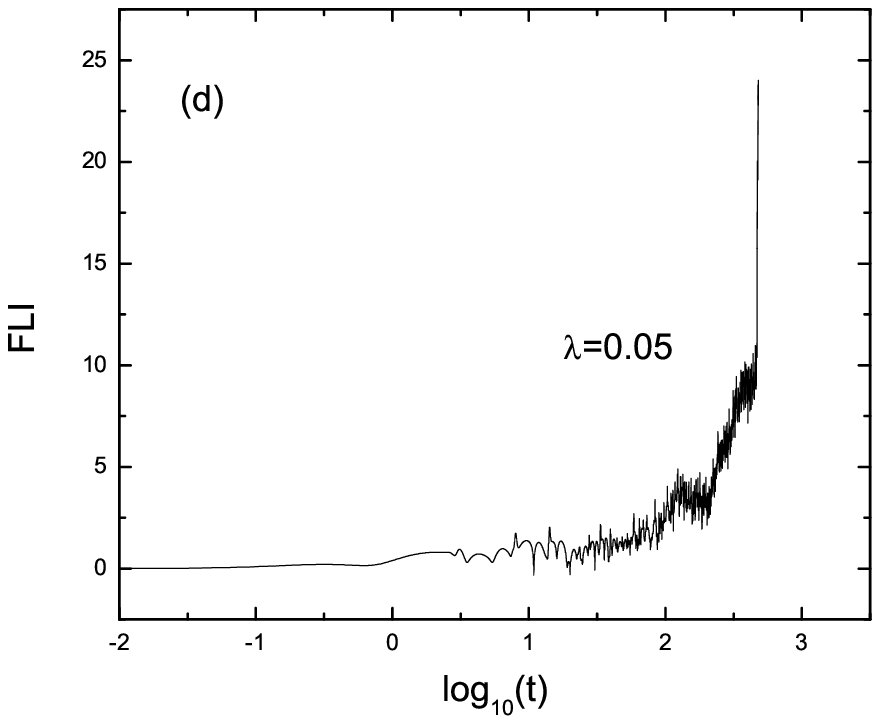}}
\caption{\label{FLIEPS}The Fast Lyapunov indicator with two nearby trajectories for $H=0$, $r_{0}=5$, $p_{r0}=2.670855$ and $\theta_{0}=0$, which is the same initial conditions as that in FIG.\ref{PS}.}}
\end{figure}

Froeschle and Lega \cite{CMDA} used the magnitude of tangent vectors to compute the FLI.
Wu et. al.\cite{PLAWu,APJWu,PRDWu} proposed an improved version of FLI with two nearby trajectories is described as
\begin{eqnarray}
\label{metricfv3}
FLI(t)= \log_{10}\frac{|d(t)|}{|d(0)|}.
\end{eqnarray}
In order to avoid the two orbits expand too fast, a renormalization procedure should be used in the computation of Eq.(34).
So its practical computation satisfies the following requirement
\begin{eqnarray}
\label{metricfv4}
FLI= -k[1+\log_{10}d(0)]+\log_{10}\frac{|d(t)|}{|d(0)|},
\end{eqnarray}
where $k$ stands for the sequential number of renormalization.
By using this algorithm, we could distinguish chaos with order by measuring the exponential or polynomial time rate of divergence of two nearby trajectories.
The method of FLI in FIG.\ref{FLIEPS} confirms that the orbital in FIG.\ref{PS}(a) and FIG.\ref{PS}(b) are ordered, but FIG.\ref{PS}(c) and FIG.\ref{PS}(d) are chaotic.
We find that the method of FLI is sufficient to distinguish between the ordered and chaotic orbital when the integration time only adds up to $10^{2}$.
Again, by using the FLI, we furthermore confirm that when the GB parameter is lower than the threshold, the orbit is regular,
and the GB parameter is higher than the threshold, the orbit is chaotic.

\section{Conclusions and discussion}\label{SectionC}

In this paper, we first study the dynamical behaviors of the ring string in the AdS-GB black hole
by the Poincare sections. We find that there are a threshold value for the GB parameter $\lambda$, below which the behavior of the system is regular and above which the behavior becomes gradually chaotic. As the GB parameter $\lambda$ increases, the chaotic behavior becomes stronger,
which is different from the case in the AdS-Schwarzschild black hole.
The dynamical behavior of the ring string in the AdS-Schwarzschild black hole is weakly chaotic \cite{1007.0277}, which corresponds to the case of $\lambda=0$ here.
But here, we find the strong chaotic behavior in the system of the ring string in AdS-GB black hole for large GB parameter $\lambda$.
Therefore, the chaotic dynamical behavior of the ring string in the asymptotically AdS spaces is not generic, maybe it also depend on the other bulk parameter of the black hole.
Furthermore, we confirm our findings by the Fast Lyapunov indicator.

In the future, a lot of work and problems deserve our further studies. First of all, we can extend our investigations to the asymptotically Lifshitz geometry or the geometry with hyperscaling violation \cite{0812.0530,1005.4690}, where the isotropic between the time and space is broken. Furthermore, we can compare the role that the asymptotic conditions,
the asymptotically AdS and asymptotically Lifshitz background, played.

\begin{acknowledgments}
D. Z. Ma is supported by the Natural Science Foundation of China under Grant
No. 11263003.
J. P. Wu is supported by the Natural Science Foundation of China under Grant
Nos. 11305018 and 11275208.

\end{acknowledgments}

\end{document}